\pdfoutput=1
\newcommand{\br}{\boldsymbol{r}}
\newcommand{\bk}{\boldsymbol{k}}
\newcommand{\bn}{\boldsymbol{n}}
\newcommand{\tp}{\tau_p}
\newcommand{\tm}{\tau_m}
\newcommand{\um}{u_{max}}
\newcommand{\vm}{v_{max}}
\newcommand{\sigs}{\sigma_{spin}^2}
\newcommand{\sigf}{\sigma_{spin}^4}
\newcommand{\rb}{\bar{R}}
\newcommand{\rbt}{\tilde{\bar{R}}}
\newcommand{\var}{\textnormal{var}}

\documentclass[%
superscriptaddress,
reprint,
showpacs,
 amsmath,amssymb,
 aps,
]{revtex4-1}

\usepackage{graphicx}
\usepackage{dcolumn}
\usepackage{bm}
\usepackage{epstopdf}
\usepackage{ifpdf}

\begin{document}

\title{Nanoscale Fourier-transform MRI}

\author{John M.  Nichol}

\author{Tyler R. Naibert}
\affiliation{Department of Physics, University of Illinois at Urbana-Champaign, Urbana, IL 61801}

\author{Eric R. Hemesath}

\author{Lincoln J. Lauhon}
\affiliation{Department of Materials Science and Engineering, Northwestern University, Evanston, IL 60208 }

\author{Raffi Budakian}
\email{budakian@illinois.edu}
\affiliation{Department of Physics, University of Illinois at Urbana-Champaign, Urbana, IL 61801}

\date{\today}
\begin{abstract}
We report a method for nanometer-scale pulsed nuclear magnetic resonance imaging and spectroscopy. Periodic radiofrequency pulses are used to create temporal correlations in the statistical polarization of a solid organic sample. The spin density is spatially encoded by applying a series of intense magnetic field gradient pulses generated by focusing electric current through a nanometer-scale metal constriction. We demonstrate this technique using a silicon nanowire mechanical oscillator as a magnetic resonance sensor to image $^1$H spins in a polystyrene sample. We obtain a two-dimensional projection of the sample proton density with approximately 10-nm resolution. 

\end{abstract}

\maketitle

\section{Introduction}
Nuclear magnetic resonance imaging (MRI) is a powerful technique for noninvasive three-dimensional biological and materials imaging~\cite{Ernst:1997}. In large part, MRI has proved so successful because it offers a host of sophisticated methods that can be used to image samples in a variety of informative ways. In general, all MRI techniques rely on accurate determination of magnetic resonance frequencies. For example, the locations of nuclear spins in a sample can be determined by applying an external magnetic field gradient, causing nuclear magnetic resonance (NMR) frequencies to vary in space. Nuclear spins of different types in the same sample can be distinguished because NMR frequencies are chemically specific. Even nuclear spins of the same type in different chemical environments can be distinguished (e.g., protons in fat and protons in water) via slight changes in resonance frequency as a result of the chemical shift interaction~\cite{stark:1988}. MRI can also probe local fluctuating magnetic fields through spin-relaxation weighted imaging~\cite{stark:1988}. Other well-known spectroscopic MRI techniques include functional magnetic resonance imaging~\cite{ogawa:1990}, diffusion tensor imaging~\cite{lebihan:1986}, and tomographic reconstruction~\cite{stark:1988}, to name a few.

Each of these common MRI techniques, and indeed most modern NMR spectroscopic techniques, use radiofrequency (rf) pulses to generate a component of the sample magnetization perpendicular to the external magnetic field. The coherent precession of the magnetization is measured and Fourier-transformed to yield the sample NMR spectrum. Since its discovery in 1966, pulsed Fourier-transform magnetic resonance~\cite{Ernst:1966} has revolutionized both NMR spectroscopy and MRI because it offers dramatically enhanced sensitivity over continuous-wave methods by allowing the simultaneous measurement of all spectral components~\cite{Fellgett:1958}. In acquisition schemes such as Fourier~\cite{kumar:1975,hoult:1979} or Hadamard encoding~\cite{bolinger:1988}, all components of the sample spectrum are averaged for the entire acquisition period. When detector noise is the limiting factor, these techniques significantly increase signal-to-noise ratio (SNR) in what is known as the multiplex advantage over methods that acquire each element of the spectrum sequentially. 

Because the nuclear magnetic moment is relatively weak~\cite{Ernst:1997}, however, the spatial resolution of inductive MRI remains limited to millimeter lengths scales in common practice and a few micrometers in the highest-resolution experimental instruments~\cite{ciobanu:2002}. Nonetheless, there is considerable interest in extending the resolution and sensitivity of magnetic resonance detection to enable spectroscopy and imaging on the nanometer scale. Promising work in this direction includes force-detected magnetic resonance~\cite{sidles:1995}, which has been used to perform three-dimensional imaging of single tobacco mosaic virus particles with resolution below 10 nm~\cite{degen:2009}, and nitrogen-vacancy-based magnetic resonance~\cite{degen:2008diamond,taylor:2008}, which has been used to detect proton resonance in nanometer-sized volumes~\cite{mamin:2013,staudacher:2013}. 

In spite of this remarkable progress, none of the classic pulsed magnetic resonance techniques have been applied to nanoscale systems because of two primary challenges. First, achieving high spatial resolution in nanoscale MRI generally requires intense static magnetic field gradients~\cite{degen:2009,grinolds:2013}. However, the presence of large static gradients makes uniform spin manipulation using rf pulses difficult and complicates NMR spectra. Second, pulsed magnetic resonance techniques cannot be used \textit{per se} because statistical spin fluctuations exceed the Boltzmann spin polarization in nanoscale samples~\cite{mamin:2003,mamin:2005}. When the statistical polarization dominates, the projection of the sample magnetization along any axis fluctuates randomly in time. For objects at the micrometer scale and above where the Boltzmann polarization dominates, many pulsed magnetic resonance techniques have been proposed~\cite{kempf:2003} and demonstrated~\cite{degen:2005,degen:2006,eberhardt:2007,eberhardt:2008,joss:2011} in force-detected experiments.

In the following, we present a new paradigm in force-detected magnetic resonance that overcomes both challenges to enable pulsed nuclear magnetic resonance in nanometer-size statistically polarized samples. In this proof-of-concept work, we demonstrate Fourier-transform spectroscopy and imaging with nanoscale resolution by periodically applying rf pulses to create correlations in the statistical polarization, or spin noise, of a solid organic sample. Gradient pulses for imaging are generated using ultrahigh current densities in a nanoscale metal constriction, and the spin noise correlations are recorded for a set of pulse configurations and Fourier transformed to give the spin density. A silicon nanowire oscillator is used as a magnetic resonance sensor to reconstruct a two-dimensional projection image of the proton density in a polystyrene sample with roughly 10-nm resolution. We also show that Fourier-transform imaging enhances sensitivity via the multiplex advantage for high-resolution imaging of statistically polarized samples. Most importantly, our protocol establishes a method by which all other pulsed magnetic resonance techniques can be used for nanoscale imaging and spectroscopy.

\section{Apparatus}
\begin{figure}
\includegraphics{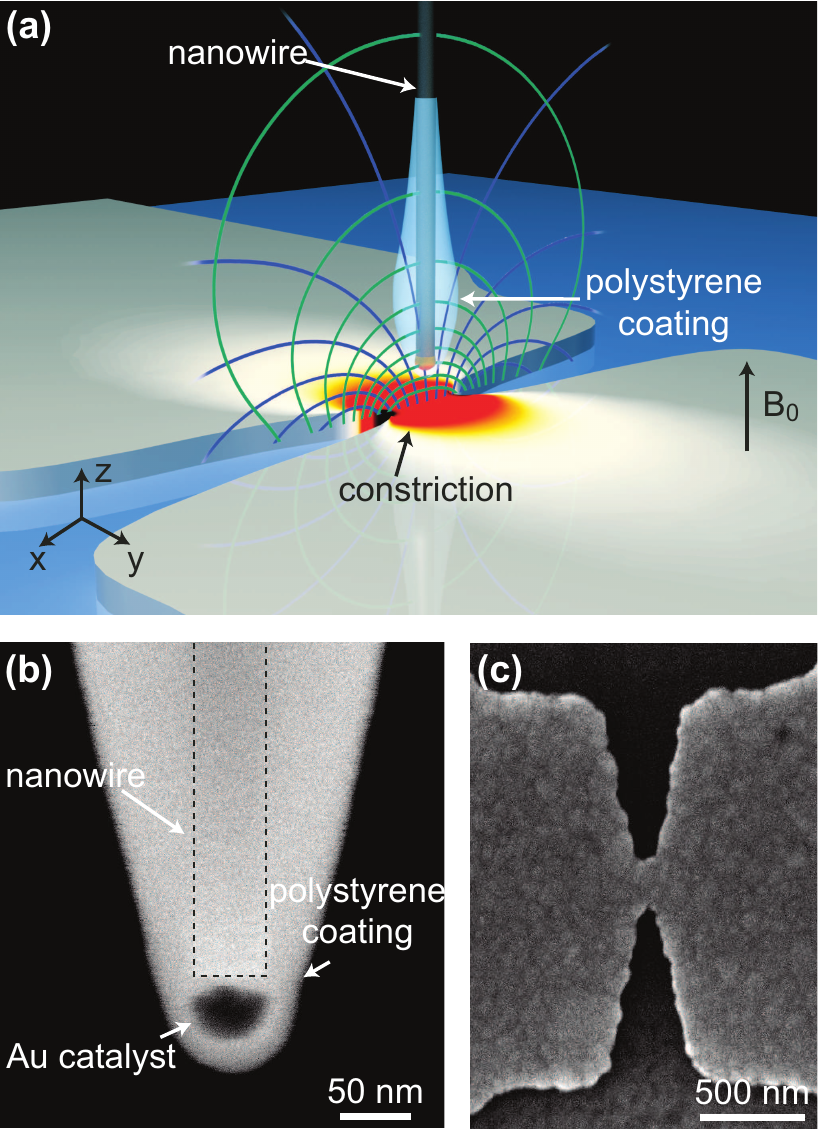}
\caption{\label{fig:apparatus} Experimental apparatus. (a) Schematic of the experimental setup. A silicon nanowire coated with polystyrene was positioned near the constriction in a Ag current-carrying wire. The locally high current density through the constriction generates intense fields and gradients used for readout, spin manipulation, and spatial encoding. During imaging, the spin density was encoded along contours of constant Larmor and Rabi frequency, which are illustrated as blue and green lines, respectively. (b) Scanning electron micrograph of a representative nanowire and polystyrene coating prepared in the same manner as the nanowire and sample used in this study. The actual nanowire and sample used here were not imaged to avoid electron damage. The dashed lines indicate the outer diameter of the nanowire. (c) Scanning electron micrograph of the constriction used in this study. The constriction is 100 nm thick and 240 nm wide.} 
\end{figure}

Figure 1(a) shows a schematic of the apparatus. A key element of the experiment is an ultrasensitive silicon nanowire force transducer~\cite{nichol:2008}, which acts as the magnetic resonance sensor. The nanowire vibrates in response to the force of interaction between the protons in the sample and the time-varying inhomogeneous magnetic field produced by a nanometer-size constriction in a current-carrying metal wire. The sample consists of a thin polystyrene coating on the tip of the silicon nanowire [Fig. \ref{fig:apparatus}(b)]. The nanowire used in this study was grown epitaxially on a Si[111] substrate using a controlled-diameter vapor-liquid-solid approach with silane as a precursor at 600 $^\circ$C~\cite{perea:2008}. The nanowire was roughly 15 $\mu$m long, with a tip diameter of 50 nm. The fundamental flexural mode had a spring constant k = 150 $\mu$N/m, a resonance frequency $\omega_0/2\pi$ = 333 kHz, and an intrinsic quality factor Q = $1.8 \times 10^4$ at a temperature of approximately 6 K. The displacement of the nanowire was measured using a polarized fiber-optic interferometer~\cite{nichol:2008,nichol:2012}. 

The current-carrying wire consists of a lithographically patterned constriction in a Ag film [Fig. \ref{fig:apparatus}(c)]. The constriction focuses current passing through the film to densities exceeding $3 \times 10^8$ Acm$^{-2}$. Such locally intense current densities generate (1) large time-dependent magnetic field gradients that couple nuclear spins in the sample to the resonant displacement of the nanowire, (2) rf magnetic fields to excite magnetic resonance in the sample, and (3) pulsed gradients for imaging. The constriction used in this study was 240 nm wide and 100 nm thick (see Supplemental Material). Both the nanowire substrate and constriction were cooled to 4.2 K in high vacuum, and the sample was positioned 40 nm above the center of the constriction. A small superconducting solenoid provided the static field $B_0 = 0.183$ T along the $z$ direction. We used the MAGGIC spin detection protocol~\cite{nichol:2012} to measure the longitudinal component of the proton statistical polarization in the sample near the constriction. In the MAGGIC protocol, an oscillating electric current through the constriction generates a magnetic field gradient that alternates at the SiNW resonance frequency. The force of interaction between the spins in the sample and the alternating inhomogeneous magnetic field induces an \r{A}ngstrom-scale vibration of the SiNW, which is measured using the optical interferometer.

\section{Spin noise encoding}
\begin{figure}
\includegraphics{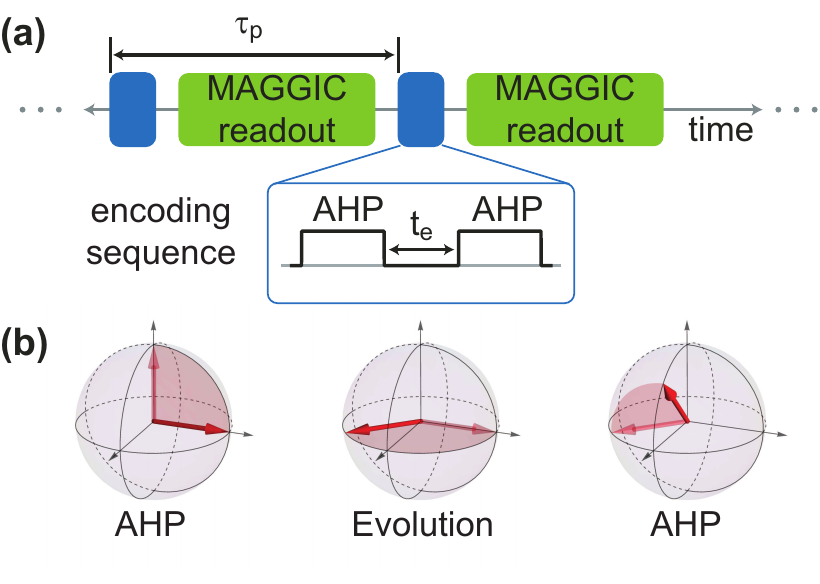}
\caption{\label{fig:fid} Spin noise encoding. (a) Periodic encoding pulses are inserted in the MAGGIC protocol every $\tau_p$. For the free-precession measurement, the sequence consists of two adiabatic half passages (AHP) separated by an evolution period $t_e$. (b) Illustration of the pulse sequence showing the evolution of a spin initially oriented along the $z$ axis.}
\end{figure}

Fourier-transform imaging and spectroscopy involve measuring the transverse component of the sample magnetization as it precesses around an external magnetic field after an rf excitation pulse. To accomplish this using the MAGGIC protocol, we used an encoding pulse sequence that is related to a Ramsey-fringe measurement~\cite{ramsey:1950}, which projects the coherent evolution of the magnetization onto the longitudinal axis. The sequence consists of an adiabatic half-passage (AHP)~\cite{garwood:2001}, an evolution period $t_e$, and a time-reversed AHP [Fig. \ref{fig:fid}(a)]. The first AHP rotates the spins away from the $z$ axis onto the $xy$ plane. During the period $t_e$, the spins precess about $B_0$. The second AHP, which is phase-shifted by $\phi (t_e)=-\gamma B_0 t_e$  relative to the first AHP, projects the magnetization back onto the $z$ axis [Fig. \ref{fig:fid} (b)]. Here $\gamma / 2\pi= 42.6$ MHz/T is the proton gryomagnetic ratio. The time-dependent phase shift creates a longitudinal projection that oscillates at the Larmor frequency as $t_e$ varies. 

Because the statistical polarization fluctuates randomly, the encoding has no effect on the mean or variance of the polarization. If, however, the sequence is inserted repeatedly (e.g. every $\tau_p$, where $1/\tau_p$ is the repetition rate) in the MAGGIC protocol, the encoding creates measurable correlations in the force signal, provided that $\tau_p \ll \tau_m$, where $\tau_m$ is the statistical spin correlation time.  In the Supplemental Material, we show that the time-averaged autocorrelation $\bar{R}_{ff}(\tau_p,t_e)$ of the force signal  at lag $\tau_p$ is
\begin{equation}
\bar{R}_{ff}(\tau_p,t_e)=\frac{e^{-\frac{\tau_p}{\tau_m}} \mu ^2 D^2}{2} \int d\bm{r} \rho (\bm{r}) G^2(\bm{r}) M(t_e,\bm{r}).
\end{equation}
Here, $\mu$ is the spin magnetic moment, $D$ is the MAGGIC gradient modulation duty cycle~\cite{nichol:2012},  $\rho(\bm{r})$ is the spin density, $G(\bm{r})$ is the gradient modulation strength, and  $M(t_e,\bm{r})$ describes the effect of the encoding. In particular,  $M(t_e,\bm{r})=1-2 P_{flip}(t_e,\boldsymbol{r})$, where $P_{flip}(t_e,\bm{r})$ is the probability for a spin located at $\bm{r}$ to reverse its orientation after a single encoding sequence. For example, if a single encoding pulse has a unit probability to invert the spin, then $M(t_e,\bm{r})=-1$; if the encoding pulse has no effect on the spin orientation, $M(t_e,\bm{r})=1$. The method presented here of measuring correlations in the spin noise is related to previous spectroscopic approaches~\cite{mamin:2005,carson:2009,leskowitz:2003,poggio:2007}, which correlate the polarization before and after an encoding pulse. 

\begin{figure}
\includegraphics{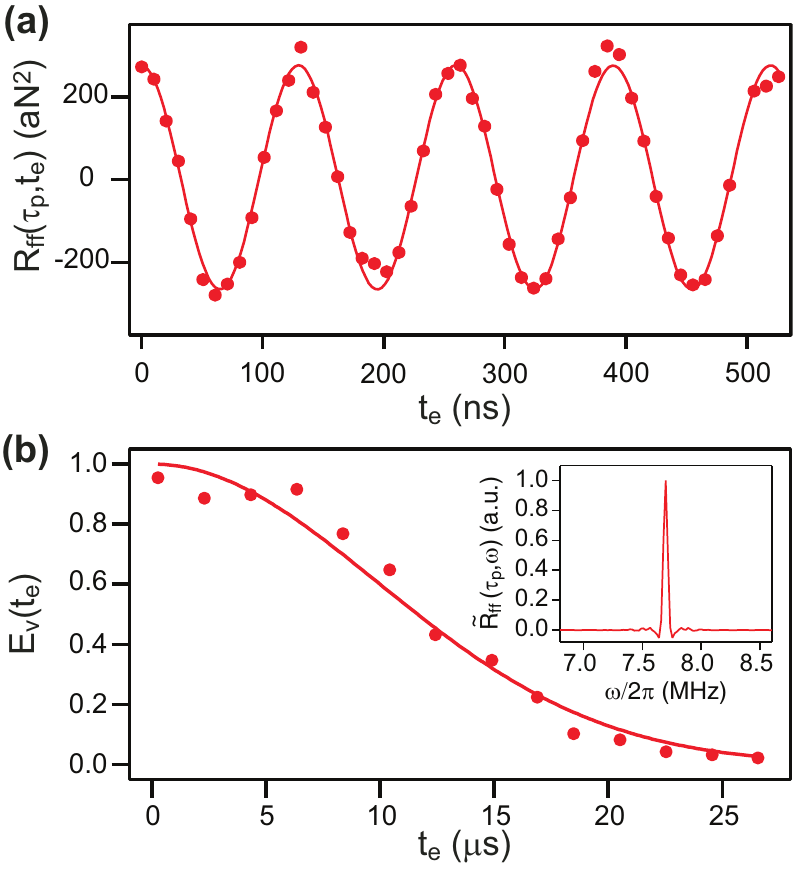}
\caption{\label{fig:fid-data} Free precession in a statistically polarized sample. (a) Autocorrelation of the force signal $R_{ff}(\tau_p,t_e)$ and fit to a cosine. (b) Amplitude of the free precession and fit to a Gaussian. From the fit, we infer that $T_2^* = 14$ $\mu$s. Inset: Proton NMR spectrum of the statistically polarized sample. }
\end{figure}

For the encoding sequence described above, $M(t_e,\bm{r})= E_v (t_e)\cos(\gamma B_0 t_e)$. The free precession will decay with an envelope $E_v(t_e)$ due to the fluctuating local fields experienced by the spins. To verify the encoding procedure, we measured the Larmor precession of the statistical polarization by sweeping $t_e$ [Fig. \ref{fig:fid-data}(a)]. The decay envelope is well-described by a Gaussian: $E_v(t_e)=e^{-(t_e/T_2^*)^2}$ with  $T_2^* = 14$ $\mu$s [Fig. \ref{fig:fid-data}(b)], consistent with previous measurements in polystyrene~\cite{froix:1976}. By cosine-transforming the data, we obtain the nuclear magnetic resonance spectrum of our statistically polarized sample [Fig. \ref{fig:fid-data}(b)].

\section{Fourier-transform imaging}
The essential feature of the free-precession encoding is the use of repeated, identical pulse sequences to induce correlations in the spin noise. Such a paradigm permits the use of established pulsed magnetic resonance techniques not only for spectroscopy, but also for imaging of statistically polarized samples. Fourier encoding, for example, uses a pulsed gradient during the free precession to encode the location of a spin in the phase or frequency of its Larmor precession. With the use of the constriction, which enables the generation of pulsed gradients, this technique can be adapted for nanoscale imaging.	

Static current through the constriction produces a strong gradient in the $x$ direction of the total field $B_{tot}(\boldsymbol{r})$. Additionally, rf current through the constriction at frequency $\gamma B_0$ produces a field in the rotating frame $B_1(\bm{r})=B_x(\bm{r})/2$, which varies strongly in the $z$ direction. These two independent gradients enable spatial encoding in two dimensions [Fig. \ref{fig:apparatus}(a)]. To simplify notation, we set $u \equiv \omega_{Rabi}(\bm{r})= \gamma B_1 (\bm{r})$ and $v \equiv \omega_{Larmor}(\bm{r})= \gamma B_0 (\bm{r})$. Because neither $u$ nor $v$ vary appreciably with respect to $y$ over the dimensions of the sample for fixed $x$ and $z$, we make the reasonable assumptions that both $u$ and $v$ are independent of $y$ for the purposes of imaging, i.e., $u(\bm{r})= u(x,z)$ and $v(\bm{r})=v(x,z)$. 

\begin{figure*}
\includegraphics{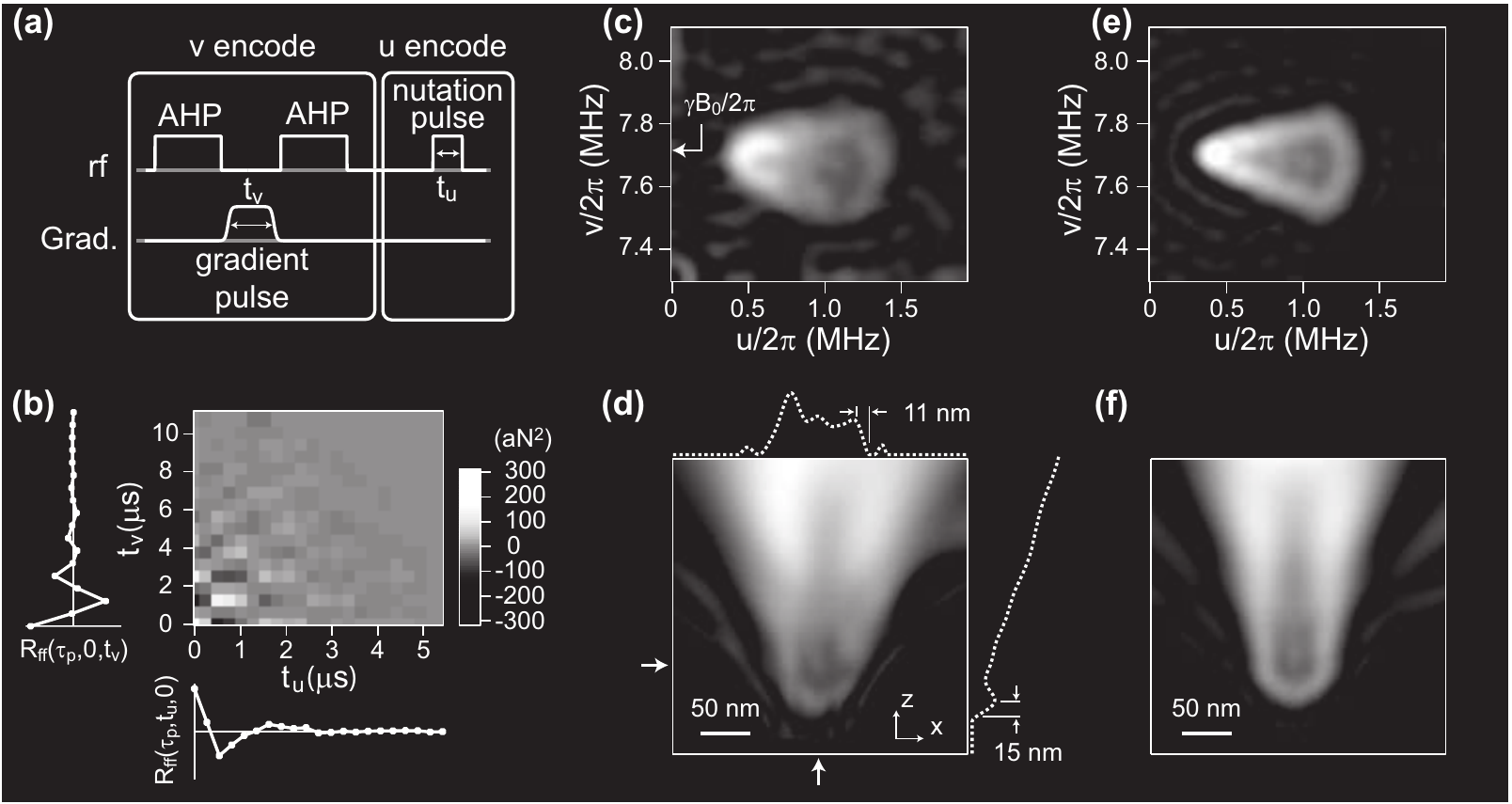}
\caption{\label{fig:images}Two-dimensional MRI of the polystyrene sample. (a) Image encoding sequence. In the  $v$-encoding step, the spins precess in the presence of a gradient for a time $t_v$. In the $u$-encoding step, the spins precess about the effective field in the rotating frame for a time $t_u$. (b) Raw data. Cross-sections corresponding to $R_{ff}(\tau_p,0,t_v)$ and $R_{ff}(\tau_p,t_u,0)$ are shown. (c) Signal density in the $(u,v)$ coordinate system obtained by cosine-transforming the raw data. The arrow indicates the position of $\gamma B_0/2 \pi$. (d) Real-space reconstruction of the projected spin density. The nanowire and gold catalyst are clearly visible through the polystyrene in the image as a reduction in the spin density. The cross-sections above and to the right of the image are taken along the lines indicated by the arrows. (e) Simulated signal density in $(u,v)$-space calculated for the sample and nanowire geometry shown in Fig. \ref{fig:apparatus}(b). (f) Real-space reconstruction of the simulation in (e). }
\end{figure*}

To encode the spin density along  $v$-contours, a pulse sequence similar to the free-precession sequence is used, except that a static gradient pulse of length $t_v$ is applied during the evolution period [Fig. \ref{fig:images}(a)]. To encode along $u$-contours, an rf pulse of length $t_u$ with center frequency $\gamma B_0$ is used to nutate spins about the effective field in the rotating frame by an angle $\gamma B_1 (\bm{r}) t_u$.  By incrementing the gradient pulse lengths, we record the Fourier transform of the two-dimensional projection of the spin density.

For the sequence discussed above, $M(t_u,t_v,\bm{r})= E_u(t_u) E_v(t_v)\cos(u(\bm{r})t_u)\cos(v(\bm{r})t_v)$, where $E_u(t_u)$ describes the transverse spin relaxation in the rotating frame \footnote{$E_u(t_u)$ could not be measured in the present experiment because the constriction produces a highly inhomogeneous rf field}.  Hence, 
\begin{eqnarray}
\bar{R}_{ff}(\tau_p,t_u,t_v)=\frac{e^{-\frac{\tau_p}{\tau_m}} E_u(t_u) E_v(t_v) \mu ^2 D^2}{2}\\ \nonumber \times \int dudv \, p(u,v) \cos(u t_u) \cos(v t_v).
\end{eqnarray}
where $p(u,v)=G^2(u,v)J(u,v)\int dy \rho (y,u,v)$ is the projected signal density in the $(u,v)$ coordinate system, and $J(u,v)$ is the Jacobian of the $(x,z) \rightarrow (u,v)$ coordinate transformation. We have also assumed that the gradient is independent of $y$ for fixed $x$ and $z$ in the sample, i.e., $G(\bm{r})=dB_z(x,z)/dx$.

To record an image, $R_{ff}(\tau_p,t_u,t_v)$ was measured for 305 different $(t_u,t_v)$ configurations [Fig. \ref{fig:images}(b)]. The data were cosine-transformed to obtain the frequency-space projection of the proton density [Fig. \ref{fig:images}(c)] and the real-space representation of the proton density [Fig. \ref{fig:images}(d)] (see Supplemental Material). The reconstructed spin density strongly resembles the expected shape of the polystyrene coating [Fig. \ref{fig:apparatus}(b)]. The nanowire and the gold catalyst particle are clearly visible through the polystyrene in the image as a reduction in the spin density. Figures \ref{fig:images}(e) and \ref{fig:images}(f) are simulations based on the expected shape of the sample [Fig. \ref{fig:apparatus}(b)]. Both simulations appear qualitatively similar to the actual data and image. 

To determine the spatial resolution in the image, we simulated the image from a point source located near the tip of the sample and find the resolution in the $x$ and $z$ directions to be approximately 10 nm and 15 nm, respectively. The spatial resolution is best at the tip of the sample where the gradients from the constriction are the largest (see Supplemental Material). The maximum imaging gradients in this study were $2.0 \times 10^5$ Tm$^{-1}$ in $z$ for 56 mA of current through the constriction and $1.4 \times 10^5$ Tm$^{-1}$ in $x$ for 20 mA of current through the constriction. The $v$ encoding pulse was limited to 20 mA to avoid artifacts from the strong transverse field produced by the constriction during the gradient pulse (22 mT at the tip of the sample for 20 mA). In the future, such artifacts can be avoided by increasing $B_0$. Nonetheless, these gradients are more than $10^4$ times stronger than the highest gradients used in inductively-detected MRI~\cite{lee:2001}.  During the readout, the peak gradient was approximately $5.0 \times 10^5$ Tm$^{-1}$ for 71 mA (limited by our amplifiers) of current through the constriction, corresponding to a current density of $3.0 \times 10^8$ Acm$^{-2}$. 

While the primary aim of the present study is to demonstrate a new technique for nanoscale pulsed magnetic resonance, the spatial resolution achieved here is comparable to the best resolution obtained in nanoscale MRI~\cite{degen:2009}. Here, the resolution could be improved by working at higher static magnetic field strengths, which would enable stronger $v$ encoding pulses. In addition, the readout gradient could likely be increased by several times, which would potentially improve the SNR and resolution by a considerable amount. We have successfully tested several smaller constrictions that we have fabricated at current densities larger than $10^9$ Acm$^{-2}$ without failure. We typically observe, however, that nanowires experience a small increase in force noise during operation of the constriction. The origin of this excess force noise is not totally clear but appears to be electrostatic in nature. To achieve the best SNR, this excess noise should be minimized to enable working at high current densities.

\section{Sensitivity}
When the measurement is dominated by detector noise, multiplexed techniques, such as Fourier-transform methods, greatly enhance sensitivity by allowing the simultaneous acquisition of multiple spectral components. In general, such methods improve sensitivity by a factor of $\sqrt{N}$ over sequential-point methods, where $N$ is the number of image points. This sensitivity boost is known as the multiplex advantage~\cite{Fellgett:1958}. For inductively-detected MRI of thermally polarized samples, the dominant noise source is voltage noise from the receiver circuitry, and Fourier-encoding is commonly employed as an efficient method of imaging~\cite{brunner:1979}.  The multiplex advantage has also been exploited for force-detected NMR and MRI of thermally polarized samples~\cite{degen:2005,degen:2006,eberhardt:2007,eberhardt:2008,joss:2011}, where force noise dominates.

If, however, the object of interest is statistically polarized, as is the case for nanometer-size samples, spin noise contributes to the total noise~\cite{degen:2007}, and a new analysis is required. Below, we argue that for high-resolution imaging of statistically polarized samples, the detector noise effectively dominates the total noise for small voxel sizes. In this case, the multiplex advantage still holds, and Fourier-transform techniques offer a significant sensitivity boost for nanoscale MRI.

In the Supplemental Material we show that, in $d$ dimensions, the average SNR of an image acquired via Fourier encoding is 
\begin{equation}
SNR=e^{-\frac{\tau_p}{\tau_m}}\left( \frac{2^d N\bar{A}\tau_p}{T} + \frac{2^d N S_F}{T \sigma_{spin}^{2}} + \frac{2^d N  S_F^2}{4 T \tau_p \sigma_{spin}^4} \right)^{-\frac{1}{2}} \label{eq:snrfourier}.
\end{equation}
Here, $\sigma_{spin}^{2}$ is the variance of the spin-component of the force signal from the entire sample, $S_F$ is the oscillator force noise power spectral density, $T$ is the averaging time per point, and  $\bar{A}$, which is approximately 2 for the present experiment, characterizes the average error in the autocorrelation integrated over the sample. Spin relaxation effects have been neglected in the above estimate. For comparison, the average SNR of an image in which each voxel is measured sequentially is 	
\begin{equation}
SNR_{point}= \left( \frac{2 \tau_m }{T} + \frac{2S_F^2 N^2}{T \tau_m \sigma_{spin}^{4}} +\frac{2 S_F N}{T \sigma_{spin}^{2}}\right)^{-\frac{1}{2}}\label{eq:snrpoint}.
\end{equation}
Equation (\ref{eq:snrpoint}) assumes that the signal in each voxel is the same (see Supplemental Material).

In both Eqs. (\ref{eq:snrfourier}) and (\ref{eq:snrpoint}), the first term in the parentheses represents the spin noise, the second term represents the oscillator force noise, and the last term is the covariance of the force noise and the spin noise. In Fourier-encoding, the force noise contribution scales more favorably with $N$ than in sequential-point imaging because all voxels in the image are measured $N$ times, compared with only once in the sequential-point case. This sensitivity enhancement exemplifies the multiplex advantage when detector noise dominates. The spin noise contribution, however, scales less favorably with $N$ in Fourier-encoding, because spin noise from the entire sample contributes to every data point. 

When the number of image points is large enough such that the force noise significantly exceeds the spin noise per voxel, i.e., when $NS_F/2 \tau_m \sigma_{spin}^2 \gg 1$, then $SNR_{point} \propto 1/N$, in contrast to Fourier encoding, where $SNR \propto 1/\sqrt{N}$. In this regime, the detector noise (i.e., force noise) effectively dominates, and Fourier encoding can be expected to offer better sensitivity. The present experiment, for example, benefited from multiplexing, because $\sigma_{spin}^2 \approx 300$ aN$^2$, $S_F \approx 10$ aN$^2$Hz$^{-1}$, and $NS_F/2 \tau_m \sigma_{spin}^2 \approx 13$.

\section{Conclusion}
In this work, we have demonstrated nanoscale pulsed Fourier-transform magnetic resonance spectroscopy and imaging. Our technique relies on creating correlations in the spin noise of a nanoscale sample using rf and gradient pulses generated by a metal constriction. We have also argued that our technique provides a sensitivity enhancement for high-resolution nanoscale imaging via the multiplex advantage. 

We conclude by noting several possible extensions of our work. First, our technique could be readily extended to enable full three-dimensional encoding with constrictions capable of producing two orthogonal static gradients. A small coil could also be used to generate a uniform rf field in the sample, which would enable the use of solid-state decoupling sequences, such as the magic sandwich~\cite{degen:2006}. These pulse sequences could be used for high-resolution spectroscopy and would permit longer encoding times and better spatial resolution in imaging. Such a coil, together with gradient pulses from the constriction, could also be employed to perform nanoscale tomography. More generally, our approach serves as a model for leveraging these and other sophisticated pulsed magnetic resonance tools to aid nanoscale MRI in its progress toward atomic-scale imaging.

\begin{acknowledgments}
This work was supported by the Army Research Office through grant No. W911NF 12 1 0341 and by the Department of Physics and the Frederick Seitz Materials Research Laboratory Central Facilities at the University of Illinois. Work at Northwestern University was supported by the National Science Foundation Grant Nos. DMI-0507053 (E.R.H) and DMR-1006069 (L.J.L.).
\end{acknowledgments}

\bibliography{ftmri-arxiv}
\onecolumngrid

\setcounter{equation}{0}
\makeatletter 
\renewcommand{\theequation}{S\@arabic\c@equation} 
\newpage
\section*{Supplemental Material for\\ Nanoscale Fourier-transform MRI}
\section*{Fabrication of the constriction}
The constriction was fabricated on a single-crystal MgO (100) substrate using a liftoff process with electron beam lithography and an MMA/PMMA resist bilayer. AquaSAVE (Mitsubishi Rayon Co., Ltd.) was used as a conductive layer on top of the resist. Ag was deposited via ultra-high vacuum electron beam evaporation. The MgO surface, because it is hygroscopic, was cleaned briefly by argon ion milling \textit{in situ} before deposition. 

Conventional photoresist delaminated during aqueous development, so the contact pads and leads were defined using deep ultraviolet optical lithography and an MMA/PMMA resist bilayer. The same deposition and liftoff procedure used for the constriction was used for the pads and leads. Thin copper wires were gap welded to the pads for good electrical contact. The substrate was diced and polished to ensure that the constriction was within 30 $\mu$m of the substrate edge to allow an unobstructed optical path between the nanowire and the optical fiber. 

\section*{Derivation of the time-averaged autocorrelation}
We derive Eq. (1) in the main text.  Consider a set of  nuclear spins constituting the sample, which evolve in time under the influence of statistical fluctuations and encoding pulses. The output of our lock-in amplifier is the (time-dependent) rms amplitude of the force exerted on the nanowire by the spins:
\begin{equation}
f(t)=\frac{D}{\sqrt{2}}\sum\limits_{i=1}^{N_{spins}}\mu_i G_i m_i(t).
\end{equation}
Here, $\mu_i$ is the spin magnetic moment, $G_i$ is the peak gradient experienced by the $i^{th}$ spin, $D$ is the duty cycle of the MAGGIC readout~\cite{nichol:2012}, and $m_i(t)$ is a variable taking on the values $\pm 1$ describing the orientation of the $i^{th}$ spin as it evolves in time. 

The time-averaged (non-normalized) autocorrelation of the force signal at lag $\tau_p$ is
\begin{eqnarray}
\bar{R}_{ff}(\tau_p) &=& \lim\limits_{T \rightarrow \infty} \frac{1}{T} \int_{0}^{T} dt f(t) f(t-\tau_p)\\
&=& \lim\limits_{T \rightarrow \infty} \frac{D^2}{2T} \int_{0}^{T} dt \sum\limits_{i=1}^{N_{spins}} \sum\limits_{j=1}^{N_{spins}} \mu^2 G_i G_j m_i(t) m_j(t-\tau_p)\\
&=& \lim\limits_{T \rightarrow \infty} \frac{D^2}{2}\sum\limits_{i=1}^{N_{spins}} \sum\limits_{j=1}^{N_{spins}}   \mu^2 G_i G_j \frac{1}{T} \int_{0}^{T} dt  m_i(t) m_j(t-\tau_p)\\
&=& \lim\limits_{T \rightarrow \infty} \frac{D^2}{2}\sum\limits_{i=1}^{N_{spins}} \sum\limits_{j=1}^{N_{spins}} \mu^2 G_i G_j \delta_{i,j} \frac{1}{T} \int_{0}^{T} dt   m_i(t) m_i(t-\tau_p)\\
&=& \frac{D^2}{2}\sum\limits_{i=1}^{N_{spins}} \mu^2 G_i^2 \bar{R}_{mm,i}(\tau_p) \\
&\approx& \mu^2 \frac{D^2}{2} \int d\br \rho(\br) G^2(\br) \bar{R}_{mm}(\tau_p,\br).
\end{eqnarray}
We have made use of the fact that statistical flips between different spins are independent in the $4^{\textnormal{th}}$ equality, and in the last equality, we have passed into the continuum limit, where $\rho(\br)$ is the spin number density, and $G(\br)$ is the peak value of the gradient. Note that
\begin{eqnarray}
\bar{R}_{mm}(\tp,\br)&=& \lim\limits_{T \rightarrow \infty} \frac{1}{T} \int_{0}^{T} dt \, m(\br,t)m(\br,t-\tp)\\
&=&\langle m(\br,t) m(\br,t-\tp) \rangle\\
&=&P_+ (\br) - P_-(\br),
\end{eqnarray}
where $P_+(\br)$ is the probability that $m(\br,t)=+m(\br,t+\tp)$, and $P_-(\br)$ is the probability that $m(\br,t)=-m(\br,t+\tp)$. Note also that 
\begin{equation}
P_+(\br)-P_-(\br)=P_{even}P_{no \, flip}(\br)+P_{odd}P_{flip}(\br)-P_{odd}P_{no \, flip}(\br)-P_{even}P_{flip}(\br),
\end{equation}
where $P_{even}$ is the probability for an even number of statistical flips during $\tp$, $P_{odd}$ is the probability for an odd number of statistical flips, $P_{flip}(\br)$ is the probability for the spin located at $\br$ to have reversed its orientation after an encoding sequence, and $P_{no \, flip}(\br)$ is the probability for no change in orientation after an encoding sequence. Since $P_{no \, flip}(\br)= 1- P_{flip}(\br)$, 
\begin{equation}
P_+(\br)-P_-(\br)=(P_{even}-P_{odd})(1-2P_{flip}(\br)).
\end{equation}
Assuming that the statistical fluctuations obey Poisson statistics~\cite{davenport:1958}, $P_{even}=(1+e^{-\tp/\tau_m})/2$, and $P_{odd}=(1-e^{-\tp/\tau_m})/2$. Hence,
\begin{equation}
\bar{R}_{mm}(\tp,\br)=e^{-\tp/\tm}M(\br),
\end{equation}
where $M(\br)=1-2P_{flip}(\br)$ describes the effect of a single encoding pulse on the spin at $\br$. $M(\br)$ depends on any parameters that affect $P_{flip}(\br)$, such as the pulse length or amplitude, for example.

\section*{Image reconstruction}
Here we describe the mathematical image reconstruction procedure for a two-dimensional data set. The generalization to other dimensions is straightforward. Recall from the main text that $\bar{R}_{ff}(\tp,t_u,t_v)=\frac{e^{-\tp/\tm}\mu^2 D^2}{2}\int du dv \, p(u,v) \cos(u t_u) \cos(v t_v)$, where $p(u,v)=G^2(u,v) J(u,v) \int dy \rho(y,u,v)$ is the projected signal density in the $(u,v)$ coordinate system, and $J(u,v)$ is the Jacobian of the $(x,z)\rightarrow (u,v)$ coordinate transformation. (We have here neglected the effects of relaxation, which will be considered in a later section.) By incrementing the pulse lengths $t_u=k_u \Delta t_u$, with $k_u=0,1,\dots , N_u-1$, and $t_v=k_v \Delta t_v$, with $k_v=0,1,\dots , N_v-1$, we may record a series $\bar{R}_{\bk}=\bar{R}_{ff}(\tp,k_u \Delta t_u, k_v \Delta t_v)$, where $\bk = (k_u,k_v)$. A discrete cosine transformation (DCT) may be applied to the data to recover the spin density. Setting $\um=\pi/\Delta t_u$ and $\vm=\pi/\Delta t_v$, the appropriate transformation is the DCT-I~\cite{wang:1983}:
\begin{equation}
\tilde{\bar{R}}_{\bn}=\frac{2}{\um}\frac{2}{\vm}\sum\limits_{k_u=0}^{N_u-1}\sum\limits_{k_v=0}^{N_v-1}\bar{R}_{\bk} w(\bk) \cos \left( \frac{\pi n_u k_u}{N_u-1}\right)  \cos \left(\frac{\pi n_v k_v}{N_v-1}\right) \label{eq:inverse},
\end{equation}
where $\bn=(n_u,n_v)$, and the weighting function is 
\begin{equation}
w(\bk)=\left\{ \begin{array}{c} 1/2\textnormal{ if } k_u=0 \textnormal{ or }N_u-1\\ 1 \textnormal{ otherwise } \end{array} \right\} \times \left\{ \begin{array}{c} 1/2\textnormal{ if } k_v=0 \textnormal{ or }N_v-1\\ 1 \textnormal{ otherwise } \end{array} \right\}.
\end{equation}
As $N_u,N_v \to \infty$, $\tilde{\bar{R}}_{\bn} \to \frac{e^{-\tp/\tm}\mu^2 D^2}{2}p \left( \frac{n_u}{N_u-1}\um, \frac{n_v}{N_v-1}\vm\right)$.

In the experiment, the pulse interval was $\tp=11$ ms, and the spin correlation time was $\tm \approx 400$ ms. To record an image, $R_{ff}(\tp,t_u,t_v)$ was measured for 305 different $(t_u,t_v)$ configurations and averaged for 29 minutes per point. The $u$ encoding pulse was stepped in increments of 0.26 $\mu$s up to 5.2 $\mu$s, and the $v$ encoding pulse was stepped in increments of 0.625 $\mu$s up to 10.6 $\mu$s. The data were zero-padded by a factor of 4 in each dimension. The DCT-I was used to recover the signal density $p(u,v)$. To suppress noise, all negative values of $p(u,v)$ were set to zero. To obtain the $(x,z)$-space representation of the image, $p(u,v)$ was divided by $G^2(u,v)J(u,v)$, and the coordinates were transformed from $(u,v)$ to $(x,z)$. The magnetic field distribution from the constriction was calculated using COMSOL Multiphysics (COMSOL, Inc.) and was used to obtain $G(u,v)$, $J(u,v)$, and the $(u,v) \leftrightarrow (x,z)$ coordinate transformation.

To generate the simulations,  $\bar{R}_{ff}(\tp,t_u,t_v)$ was calculated for each $(t_u,t_v)$ point using the calculated magnetic field distribution and the profile of a nanowire tip and polystyrene coating [Fig. 1(b) in the main text] prepared in the same fashion as the tip and sample used in this study. The actual sample used here was not imaged in a scanning electron microscope to avoid electron beam damage.

\section*{Spatial resolution}
In the image, the voxel size increases with distance from the constriction (Fig. \ref{fig:voxels}) because the magnetic field gradients are strongest near the constriction. In $(u,v)$-space, however, the voxel dimensions are constant: $\Delta u = (2 T_u^{-1})$ and $\Delta v = (2 T_v^{-1})$, where $T_u$ and $T_v$ are the maximum $u$ and $v$ evolution times during the encoding. Because the constriction generates non-uniform gradients, the spatial resolution varies in real space. Neglecting the effects of spin relaxation, the minimum voxel dimensions in real space can be estimated as $\Delta x \approx (2 T_v v_{x,max})^{-1}$ and $\Delta z \approx (2 T_u u_{z,max})^{-1}$. Here $v_{x,max} 2\pi/\gamma$ and $u_{z,max} 2 \pi/\gamma$ are the maximum imaging gradients experienced by the sample, which were $2.0 \times 10^5$ Tm$^{-1}$ in $z$ for 56 mA of current through the constriction and $1.4 \times 10^5$ Tm$^{-1}$ in $x$ for 20 mA of current through the constriction.

\setcounter{figure}{0}
\makeatletter 
\renewcommand{\thefigure}{S\@arabic\c@figure} 
\begin{figure}
\includegraphics{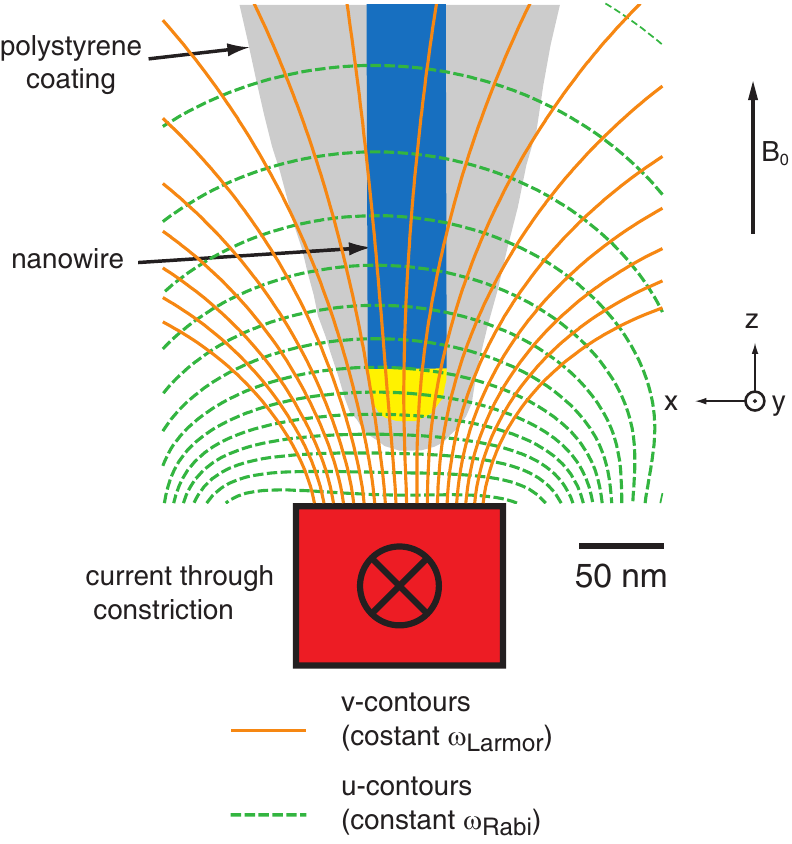}
\caption{\label{fig:voxels} Schematic illustrating the dependence of the voxel size on distance from constriction. The field contours displayed are the actual contours used for imaging. The voxel size is the smallest and the spatial resolution is the best at the tip of the sample. The magnetic field gradients used for imaging are strongest at the point in the sample closest to the constriction.  }
\end{figure} 

\section*{Error in measurement of the autocorrelation}
In practice, the measured value of the autocorrelation is the expected value plus random noise: $R_{\bk}=\bar{R}_{\bk}+r_{\bk}$. Noise in the raw data produces noise in the image: $\tilde{R}_{\bn}=\rbt_{\bn}+\tilde{r}_{\bn}$. Here we compute the expected value of the noise in the data $r_{\bk}$, and in the next section we will compute the average value of the noise in the image $\tilde{r}_{\bn}$. We assume that the force signal $f(t)$ is digitized much more rapidly that $\tp$ and that all samples between consecutive pulses are averaged together. The signal is thus filtered synchronously with the encoding using a convolution filer~\cite{degen:2007} with time constant $\tp$. Hence a continuous record of the measured signal $s_{\bk}(t)=f_{\bk}(t) + n(t)$ containing the desired force signal $f_{\bk}(t)$ and random instrumentation noise $n(t)$ spanning a time $T$ becomes a discrete set of points $s_{\bk,i}(t)=f_{\bk,i} + n_i$, where $i=0,1, \dots, N_{pts}-1$, and $N_{pts}=T/\tp$. We have retained the index $\bk$ to indicate the encoding pulse configuration, and the index $i$ is associated with a point in time. The autocorrelation is
\begin{eqnarray}
R_{\bk}&=&\frac{1}{N_{pts}}\sum \limits_{i=0}^{N_{pts}-1}s_{\bk,i} s_{\bk,i+1}\\
&=& \frac{1}{N_{pts}}\sum \limits_{i=0}^{N_{pts}-1} f_{\bk,i} f_{\bk,i+1}+n_{i} n_{i+1}+f_{\bk,i} n_{i+1}+n_{i} f_{\bk,i+1}.
\end{eqnarray}

Only the first term $\frac{1}{N_{pts}}\sum \limits_{i=0}^{N_{pts}-1} f_{\bk,i} f_{\bk,i+1}$ has a non-vanishing average value. However, all of the terms have a non-zero variance. It can be shown, using the results of Refs.~\cite{degen:2007,bartlett:1935,bartlett:1946} that
\begin{eqnarray}
\var\left( \frac{1}{N_{pts}} \sum \limits_{i=0}^{N_{pts}-1} n_{i} f_{\bk,i+1} \right) &=& \frac{\var(n_i)\var(f_{\bk,i})}{N_{pts}}\\
&=&\frac{(S_F/2\tp)\sigma^2_{spin}}{T/\tp}\\
&=&\frac{S_F\sigma^2_{spin}}{2T},
\end{eqnarray}
\begin{eqnarray}
\var\left( \frac{1}{N_{pts}} \sum \limits_{i=0}^{N_{pts}-1} n_{i} n_{i+1} \right) &=& \frac{\var(n_i)^2}{N_{pts}}\\
&=&\frac{(S_F/2\tp)^2}{T/\tp}\\
&=&\frac{S_F^2}{4T\tp},
\end{eqnarray}
and 
\begin{eqnarray}
\var\left( \frac{1}{N_{pts}} \sum \limits_{i=0}^{N_{pts}-1} f_{\bk,i} f_{\bk,i+1} \right) &=& \frac{\var(f_{\bk,i})^2\tp}{T} \sum \limits_{j=-\infty}^{\infty}\bar{\rho}_{\bk,j}^2+\bar{\rho}_{\bk,j-1}\bar{\rho}_{\bk,j+1}\\
&=&\frac{\sigma_{spin}^4 \tp}{T}\sum \limits_{j=-\infty}^{\infty}\bar{\rho}_{\bk,j}^2+\bar{\rho}_{\bk,j-1}\bar{\rho}_{\bk,j+1}\\
&=&A_{\bk}\frac{\sigma_{spin}^4 \tp}{T},
\end{eqnarray}
where $\var(\cdots)$ indicates the variance of the quantity in parentheses,
\begin{equation}
\sigma_{spin}^2= \frac{\mu^2 D^2}{2}\int d\br \rho(\br) G^2(\br),
\end{equation}
 is the variance of the spin component of the force signal, and 
\begin{eqnarray}
\bar{\rho}_{\bk,j}&=& \rb_{ff}(j\tp,k_u \Delta t_u, k_v \Delta t_v)/\sigma_{spin}^2\\
&=& \frac{\int_{0}^{\vm} dv \int_{0}^{\um} du \, p(u,v) \cos^j \left( \frac{\pi k_u u}{\um} \right) \cos^j \left( \frac{\pi k_v v}{\vm}\right)}{\int_{0}^{\vm} dv \int_{0}^{\um} du \, p(u,v)}
\end{eqnarray}
is the time-averaged normalized autocorrelation of the force signal at lag $j\tp$. The quantity $A_{\bk}=\sum \limits_{j=-\infty}^{\infty}\bar{\rho}_{\bk,j}^2+\bar{\rho}_{\bk,j-1}\bar{\rho}_{\bk,j+1}$ characterizes the average error in the non-normalized autocorrelation~\cite{bartlett:1946}. Putting it all together,
\begin{equation}
r_{\bk}^2=A_{\bk}\frac{\sigma_{spin}^4 \tp}{T}+\frac{S_F \sigma_{spin}^2}{T} + \frac{S_F^2}{4 T \tp}.
\end{equation}

\section*{Image signal to noise}
The orthogonality relation for the DCT-I is
\begin{eqnarray}
\sum \limits_{n_u=0}^{N_u-1}\sum \limits_{n_v=0}^{N_v-1}a^2(\bn)\cos \left( \frac{\pi k_u n_u}{N_u-1} \right)\cos \left( \frac{\pi k_u' n_u}{N_u-1} \right) \cos \left( \frac{\pi k_v n_v}{N_v-1} \right) \cos \left( \frac{\pi k_v' n_v}{N_v-1} \right) \nonumber \\
=\frac{N_u-1}{2}\frac{N_v-1}{2}\delta_{\bk,\bk'}\times \left\lbrace \begin{array}{c} 2\textnormal{ if } k_u=0 \textnormal{ or }N_u-1\\ 1 \textnormal{ otherwise } \end{array} \right\rbrace\times \left\lbrace \begin{array}{c} 2\textnormal{ if } k_v=0 \textnormal{ or }N_v-1\\ 1 \textnormal{ otherwise } \end{array} \right\rbrace, \label{eq:orthogonality}
\end{eqnarray}
where
\begin{equation}
a(\bn)=\left\lbrace\begin{array}{c} 1/\sqrt{2}\textnormal{ if } n_u=0 \textnormal{ or }N_u-1\\ 1 \textnormal{ otherwise } \end{array} \right\rbrace\times \left\lbrace\begin{array}{c} 1/\sqrt{2}\textnormal{ if } n_v=0 \textnormal{ or }N_v-1\\ 1 \textnormal{ otherwise } \end{array} \right\rbrace.
\end{equation}
Making use of Parseval's Theorem for the DCT-I,
\begin{eqnarray}
\frac{1}{(N_u-1)(N_v-1)}\sum \limits_{n_u=0}^{N_u-1}\sum \limits_{n_v=0}^{N_v-1}a^2(\bn)\tilde{r}_{\bn}^2=\frac{2}{\um^2}\frac{2}{\vm^2}\sum \limits_{k_u=0}^{N_u-1}\sum \limits_{k_v=0}^{N_v-1} w(\bk) r_{\bk}^2 \nonumber \\
=\frac{2(N_u-1)}{\um^2}\frac{2(N_v-1)}{\vm^2}\left( \bar{A} \frac{\sigma_{spin}^4 \tp}{T}+\frac{S_F \sigma_{spin}^2}{T} + \frac{S_F^2}{4 T \tp} \right). \label{eq:parseval}
\end{eqnarray}
Here $\bar{A}=\frac{1}{N_u-1}\frac{1}{N_v-1}\sum \limits_{k_u=0}^{N_u-1}\sum \limits_{k_v=0}^{N_v-1} w(\bk) A_{\bk}$. Note that $w(\bk)$ appears only to the first power to account for the normalization factor in Eq. (\ref{eq:orthogonality}). For the current experiment, we estimate that $\bar{A}\approx 2$.

A useful quantity to calculate is the average SNR in the image, which we define as the average signal divided by the root-mean-square noise. The average variance in the image is given by Eq. (\ref{eq:parseval}), and the average signal is $e^{-\tp/\tm}\sigma_{spin}^2/\um\vm$. Hence, the average SNR in $d$ dimensions is:
\begin{eqnarray}
SNR&=&e^{-\tp/\tm}(N_u-1)^{-1/2}(N_v-1)^{-1/2}\left( \frac{2^d \bar{A}\tp}{T} + \frac{2^d S_F}{T \sigma_{spin}^{2}} + \frac{2^d S_F^2}{4 T \tp \sigma_{spin}^4} \right)^{-1/2}\\
&\approx&e^{-\tp/\tm}\left( \frac{2^d N \bar{A}\tp}{T} + \frac{2^d N S_F}{T \sigma_{spin}^{2}} + \frac{2^d N  S_F^2}{4 T \tp \sigma_{spin}^4} \right)^{-1/2},
\end{eqnarray}
provided that $N_u \gg 1$ and $N_v \gg 1$, and where $N = N_uN_v$.

For comparison, the SNR of a sequential-point image may be calculated using the results of Ref.~\cite{degen:2007}. Assuming that the signal per voxel is $\sigma_{spin}^2/N$, where $N$ is the total number of points, the noise energy in each voxel is $\frac{2 \tm}{T}\left( \frac{\sigf}{N^2} + \frac{2S_F^2}{4 \tm^2} +\frac{2 S_F \sigs}{2 \tm N}\right)$. Hence,
\begin{eqnarray}
SNR_{point}&\approx& \frac{\sigs}{N} \left(\frac{2 \tm}{T}\right)^{-1/2} \left( \frac{\sigf}{N^2} + \frac{2S_F^2}{4 \tm^2} +\frac{2 S_F \sigs}{2 \tm N}\right)^{-1/2}\\
&=& \left( \frac{2 \tm }{T} + \frac{2S_F^2 N^2}{T \tm \sigf} +\frac{2 S_F N}{T \sigs}\right)^{-1/2}.
\end{eqnarray}

\section*{Spin relaxation}
We now discuss the effects of spin relaxation on spatial resolution and SNR. Expanding on Eq. (\ref{eq:inverse}) and dropping the overall factor of $e^{-\tp/\tm} \mu^2 D^2/2$:
\begin{eqnarray}
\tilde{\bar{R}}_{\bn}&=&\frac{2}{\um}\frac{2}{\vm}\sum\limits_{k_u=0}^{N_u-1}\sum\limits_{k_v=0}^{N_v-1}\bar{R}_{\bk} w(\bk) \cos \left( \frac{\pi n_u k_u}{N_u-1}\right)  \cos \left(\frac{\pi n_v k_v}{N_v-1}\right)\\
&=& \int_{0}^{\vm} dv \int_{0}^{\um} du \, p(u,v) \frac{4}{\um \vm}  \\
&\times& \sum\limits_{k_u=0}^{N_u-1}\sum\limits_{k_v=0}^{N_v-1} w(\bk) \cos \left( \frac{\pi n_u k_u}{N_u-1}\right) \cos \left( \frac{\pi n_u u}{\um}\right) \cos \left( \frac{\pi n_v k_v}{N_v-1}\right) \cos \left( \frac{\pi n_v v}{\vm} \right) \nonumber \\
&=& \int_{0}^{\vm} dv \int_{0}^{\um} du \, p(u,v) g(u,v,u',v')\\
&\approx& p(u',v'),
\end{eqnarray}
where $g(u,v,u',v')=\frac{4}{\um \vm} \sum\limits_{k_u=0}^{N_u-1}\sum\limits_{k_v=0}^{N_v-1} w(\bk) \cos \left( \frac{\pi n_u k_u}{N_u-1}\right) \cos \left( \frac{\pi n_u u}{\um}\right) \cos \left( \frac{\pi n_v k_v}{N_v-1}\right) \cos \left( \frac{\pi n_v v}{\vm} \right)$, $u' = n_u \um/(N_u-1)$, and $v' = n_v \vm/(N_v-1)$. Note that $g(u,v,u',v') \to \delta(u-u',v-v')$ as $N_u,N_v \to \infty$. Note also that $\int_{0}^{\vm} dv' \int_{0}^{\um} du' g(u,v,u',v') = 1$. The kernel $g(u,v,u',v')$ is strongly peaked about $u=u'$ and $v=v'$ and defines the impulse response of the image transformation and the resulting spatial resolution.

With regard to spatial resolution, the effect of spin relaxation is to modify the shape of the kernel: $g\to \frac{4}{\um \vm} \sum\limits_{k_u=0}^{N_u-1}\sum\limits_{k_v=0}^{N_v-1} w(\bk) E(\bk) \cos \left( \frac{\pi n_u k_u}{N_u-1}\right) \cos \left( \frac{\pi n_u u}{\um}\right) \cos \left( \frac{\pi n_v k_v}{N_v-1}\right) \cos \left( \frac{\pi n_v v}{\vm} \right)$, where $E(\bk)$ describes the effect of spin relaxation. $E(\bk)$ is expected to be of the form $E(\bk) = e^{-(k_u \Delta t_u / T_{2 \rho}^*)^2}e^{-(k_v \Delta t_v / T_{2}^*)^2}$, where $T_{2 \rho}^*$ is the transverse spin relaxation time in the rotating frame. Provided that $E(0)=0$, as is usually the case, the average value of the signal density is preserved. The form of $E(\bk)$ affects the shape of $g(u,v,u',v')$ and hence the spatial resolution. In general, $E(\bk) \to 0$ as $|\bk| \to \infty$. The more rapidly $E(\bk)$ decays, the broader $g(u,v,u',v')$ becomes. Spin relaxation affects SNR via $A_{\bk} \to A_{\bk}  E^2(\bk)$. Although the spatial resolution in the image degrades the more rapidly $E(\bk)$ decays to zero, the SNR can be expected to improve slightly because of the reduced spin noise.

\end{document}